\documentclass[fleqn,twoside]{article}
\usepackage{amssymb}
\usepackage{amsfonts}
\usepackage{espcrc2}
\usepackage[dvips]{graphicx}

\newcommand{\AmS}{{\protect\the\textfont2
  A\kern-.1667em\lower.5ex\hbox{M}\kern-.125emS}}
\begin{document}

\title{Method of Brackets and Feynman diagrams evaluation}
\author{Iv\'{a}n Gonz\'{a}lez\thanks{%
Supported by Basal Project FB0821} \\
Department of Physics and Center of Subatomic Studies\\
Universidad T\'{e}cnica Federico Santa Mar\'{\i}a \\
Casilla 110-V, Valpara\'{\i}so, Chile}
\date{}
\maketitle

\begin{abstract}
We present a new heuristic method for the evaluation of Feynman diagrams and
other definite integrals : the method of brackets. The operational rules are
described and the method is illustrated with simple examples. Also, we
demostrate that this technique is validated through a generalization of the
Ramanujan's master theorem, which somewhat explains its origin.
\end{abstract}

\section{Introduction}

\qquad The aim of this work is to present an improved and more general
technique related with NDIM (Negative Dimensional Integration Method) , \cite%
{IRi}, \cite{CAn2}, \cite{CAn1}, \cite{ASu11}, \cite{ASu13}, \cite{ASu14},
which we have called Method of Brackets. This modification to NDIM was
originally presented in \cite{11} in the context of multiloops Feynman
integrals. A complete description of the operational rules of the method,
together with a variety of examples related with Feynman diagrams and a
generalization to arbitrary integrals, was discussed in \cite{10}. In Ref.
\cite{12} the usefulness of this technique in analytic continuation of
trascendental functions was also discussed, specifically for hypergeometric
functions of the form $_{q}F_{q-1}$. The method of brackets is a versatile
and simple technique for evaluating Feynman diagrams up to a certain level
of difficulty. The reason for this is the complexity of solutions, which are
presented in terms of multiple hypergeometric series in the general case.
The method of brackets is a heuristic method for the evaluation of definite
integrals, whose great advantage is to reduce the evaluation of a large
class of definite integrals to the solution of a linear system of equations.

Until now this technique does not have a rigorous mathematical proof. In
this work we show that the method of brackets is a generalization of
Ramanujan's master theorem (RMT). However, this theorem is not sufficient to
explain mathematically the bracket's technique in complete form.

In our case the application of the method of brackets to Feynman diagrams
requires the Schwinger's parametric representation of diagram. Then through
of a systematic procedure it is possible to obtain the analytical solution
to this diagram as a sum of hypergeometric functions. In this work we
present two simple examples describing, step by step the application of the
method of brackets and the Ramanujan's master theorem.

\section{General momentum representation and Schwinger%
\'{}%
s representation}

An arbitrary diagram with $L$ loops, $N$ propagators and $E$ independent
external lines, has the following associated momentum integral in $D$
dimensions in Minkowski space,%
\begin{equation}
\begin{array}{ll}
G= & \int \frac{d^{D}q_{1}}{i\pi ^{D/2}}...\frac{d^{D}q_{L}}{i\pi ^{D/2}} \\
&  \\
& \times \frac{1}{(B_{1}^{2}-m_{1}^{2}+i0)^{\nu _{1}}}...\frac{1}{%
(B_{N}^{2}-m_{N}^{2}+i0)^{\nu _{N}}},%
\end{array}%
\end{equation}%
where we define (explicit or implicity):

\begin{itemize}
\item $B_{j}\longrightarrow $Momentum of the $j$-$th$ $\left(
j=1,...,N\right) $ propagator or internal line. It is a linear combination
of external momenta $\left\{ p\right\} $ and internal momenta $\left\{
q\right\} $.

\item $\nu _{j}\longrightarrow $Arbitrary indices $\left( j=1,...,N\right) $.

\item $p_{k}\longrightarrow $External momentun $\left( k=1,...,E\right) $.

\item $q_{k}\longrightarrow $Internal momentum $\left( k=1,...,L\right) $.

\item $m_{j}\longrightarrow $Mass associated to the - $j$-$th$ propagator.
\end{itemize}

In this case the corresponding Schwinger's parametric representation is
given by the equation

\begin{equation}
\begin{array}{c}
G=\frac{(-1)^{-\frac{LD}{2}}}{\prod\nolimits_{j=1}^{N}\Gamma (\nu _{j})}%
\int\limits_{0}^{\infty }d\overrightarrow{x}\;\frac{\exp \left(
\sum\limits_{j=1}^{N}x_{j}m_{j}^{2}\right) \exp \left( -\frac{F}{U}\right) }{%
U^{\frac{D}{2}}},%
\end{array}%
\end{equation}%
where $d\overrightarrow{x}=\prod\nolimits_{j=1}^{N}dx_{j}\;x_{j}^{\nu
_{j}-1} $ , $N_{\nu }=\nu _{1}+...+\nu _{N}$, $U$ and $F$ are polynomials $L$%
-$linear $ and $(L+1)$-$linear$ respectively in Schwinger's parameters. The
polynomials $U$ and $F$ can be evaluated using the general formula \cite%
{IGo1}

\begin{equation}
\begin{array}{l}
F=\sum\limits_{i,j=1}^{E}C_{ij}\;p_{i}.p_{j}, \\
\\
U=\left\vert
\begin{array}{ccc}
M_{11} & \cdots & M_{1L} \\
\vdots &  & \vdots \\
M_{L1} & \cdots & M_{LL}%
\end{array}%
\right\vert ,%
\end{array}%
\end{equation}%
where the coefficients $C_{ij}$ are given for the following determinant

\begin{equation}
\left\vert
\begin{tabular}{lllc}
$M_{11}$ & $\cdots $ & $M_{1L}$ & $M_{1(L+j)}$ \\
$\vdots $ &  & $\vdots $ & $\vdots $ \\
\multicolumn{1}{c}{$M_{L1}$} & \multicolumn{1}{c}{$\cdots $} &
\multicolumn{1}{c}{$M_{LL}$} & $M_{L(L+j)}$ \\
$M_{(L+i)1}$ & $\cdots $ & $M_{(L+i)L}$ & $M_{(L+i)(L+j)}$%
\end{tabular}%
\right\vert .
\end{equation}%
The $\mathbf{M}$ matrix, the parameter matrix, may be evaluated directly
from topology of the diagram.

\section{Rules in the Method of Brackets}

\qquad In the following we show the fundamental rules of this technique. The
technique of brackets transforms the parameter integral into series-like
structure called : "brackets expansion". We need only four basic rules for
obtaining such an expansion.

\subsection{Rule I : Exponential function expansion}

To expand the exponential function, we use the "usual" way, this is

\begin{equation}
\exp \left( -xA\right) =\sum\limits_{n}\frac{\left( -1\right) ^{n}}{\Gamma
\left( n+1\right) }x^{n}A^{n},
\end{equation}%
if the argument of exponential function is $\exp (xA)$, we expand in this way

\begin{equation}
\exp \left( xA\right) =\sum\limits_{n}\frac{\left( -1\right) ^{n}}{\Gamma
\left( n+1\right) }x^{n}\left( -A\right) ^{n}.
\end{equation}%
The reason for this is to associate to each expansion the factor $\phi _{n}=%
\frac{\left( -1\right) ^{n}}{\Gamma \left( n+1\right) }$ as a simple
convention.

\subsection{Rule II : Integration symbol and its equivalent bracket}

This rule corresponds to the definition of the bracket symbol. The structure
$\int x^{a_{1}+a_{2}+...+a_{n}-1}$ $dx$ is replaced by its respective
bracket representation

\begin{equation}
\int x^{a_{1}+a_{2}+...+a_{n}-1}dx=\left\langle
a_{1}+a_{2}+...+a_{n}\right\rangle .
\end{equation}

\subsection{Rule III : Polynomials expansion}

For polynomials we use the following representation in terms of series of
brackets

\begin{equation}
\begin{array}{l}
\left( A_{1}+...+A_{r}\right) ^{\pm \mu }= \\
\\
\sum\limits_{n_{1}}...\sum\limits_{n_{r}}\phi _{n_{1}}...\phi
_{n_{r}}\;\left( A_{1}\right) ^{n_{1}}...\left( A_{r}\right) ^{n_{r}} \\
\\
\times \frac{\left\langle \mp \mu +n_{1}+...+n_{r}\right\rangle }{\Gamma
\left( \mp \mu \right) }.%
\end{array}%
\end{equation}%
This rule is derived using rule $\left( I\right) $ and $\left( II\right) $
after applying the Schwinger's parametrization to this polynomial. An
adequate way for expanding repeated polynomials in the integral is described
in \cite{11}, the idea in this case is to minimize the complexity of the
solution.

\subsection{Rule IV : Finding the solution}

For the case of a generic series of brackets $J$

\begin{equation}
\begin{array}{ll}
J= & \sum\limits_{n_{1}}...\sum\limits_{n_{r}}\phi _{n_{1}}...\phi
_{n_{r}}\;\digamma (n_{1},...,n_{r}) \\
&  \\
& \times \;\left\langle a_{11}n_{1}+...+a_{1r}n_{r}+c_{1}\right\rangle ...
\\
&  \\
& \times ...\left\langle a_{r1}n_{1}+...+a_{rr}n_{r}+c_{r}\right\rangle ,%
\end{array}%
\end{equation}%
the solution is obtained using the general formula

\begin{equation}
\begin{array}{c}
J=\frac{1}{\left\vert \det \left( \mathbf{A}\right) \right\vert }\Gamma
\left( -n_{1}^{\ast }\right) ...\Gamma \left( -n_{r}^{\ast }\right) \digamma
(n_{1}^{\ast },...,n_{r}^{\ast })%
\end{array}
\label{Brackets}
\end{equation}%
where $\det \left( \mathbf{A}\right) $ is evaluated by the following
expression

\begin{equation}
\det \left( \mathbf{A}\right) =\left\vert
\begin{array}{ccc}
a_{11} & \ldots & a_{1r} \\
\vdots & \ddots & \vdots \\
a_{r1} & \cdots & a_{rr}%
\end{array}%
\right\vert ,
\end{equation}%
and $\left\{ n_{i}^{\ast }\right\} $ $\;\left( i=1,...,r\right) $ is the
solution of the linear system obtained by the vanishing of the brackets

\begin{equation}
\left\{
\begin{array}{cc}
a_{11}n_{1}+...+a_{1r}n_{r}= & -c_{1} \\
\vdots & \vdots \\
a_{r1}n_{1}+...+a_{r}n_{r}= & -c_{r}.%
\end{array}%
\right.
\end{equation}%
The value of $J$ is not defined if the matrix $\mathbf{A}$ is not invertible.

\underline{Note} : In the case where a higher dimensional series has more
summation indices than brackets, the appropriate number of free variables is
chosen among the indices. For each such choice, Rule $IV$ yields a series.
Those converging in a common region are added to evaluate the desired
integral.

In the evaluation of these formal sums, the index $n$ $\in $ $N$ will be
replaced by a number $n^{\ast }$ defined by the vanishing of the bracket.
Observe that it is possible that $n^{\ast }$ $\in $ $%
\mathbb{C}
$. For book-keeping purposes, specially in cases with many indices, we write
$\sum\limits_{n}$ instead of the usual $\sum\limits_{n=0}^{\infty }$. After
that the brackets are eliminated, those indices that remain recover their
original nature.

Some simple examples and their respective expansions in brackets:

\begin{itemize}
\item For binomial expression
\end{itemize}

\[
\begin{array}{ll}
\frac{1}{\left( A-B\right) ^{\beta }}= & \sum\limits_{n_{1}}\sum%
\limits_{n_{2}}\phi _{n_{1},n_{2}}\;A^{n_{1}}\left( -B\right) ^{n_{2}} \\
&  \\
& \times \frac{\left\langle \beta +n_{1}+n_{2}\right\rangle }{\Gamma (\beta )%
}.%
\end{array}%
\]

\begin{itemize}
\item For integral
\end{itemize}

\[
\int\limits_{0}^{\infty }dx\;\frac{x^{\alpha -1}}{\exp (Ax)}%
=\sum\limits_{n}\phi _{n}\;A^{n}\;\left\langle \alpha +n\right\rangle .
\]

\section{Ramanujan's Master Theorem (RMT)}

\qquad In the following we describe the Ramanujan's master theorem and its
relation with the method of brackets. The theorem says that for an integral $%
J=\int\limits_{0}^{\infty }dx_{1}\;x^{\nu -1}\;f\left( x\right) $, where, we
suppose that $f\left( x\right) $ admits a Taylor expansion of the form

\begin{equation}
f\left( x\right) =\sum\limits_{k}\digamma \left( k\right) \frac{\left(
-x\right) ^{k}}{k!}
\end{equation}%
in a neighborhood of $x=0$ and $f\left( 0\right) =\digamma \left( 0\right)
\neq 0$, then the solution is given by

\begin{equation}
J=\int\limits_{0}^{\infty }dx\;x^{\nu -1}f\left( x\right) =\Gamma (\nu
)\digamma \left( -\nu \right) .  \label{eq1}
\end{equation}%
This integral corresponds to the Mellin transform of $f\left( x\right) $.
The condition $\digamma \left( 0\right) \neq 0$ guarantees the convergence
of integral in $\left( \ref{eq1}\right) $ near $x=0$, for $\nu >0$. This
theorem was demostrated by Hardy \cite{Hardy}. We present a generalization
to this theorem when it is applied to the multidimensional integral

\begin{equation}
\begin{array}{ll}
J= & \int\limits_{0}^{\infty }dx_{1}\;x_{1}^{\nu
_{1}-1}\;...\int\limits_{0}^{\infty }dx_{N}\;x_{N}^{\nu _{N}-1} \\
&  \\
& \times \;f\left( x_{1},...,x_{N}\right) .%
\end{array}%
\end{equation}%
If $f\left( x_{1},...,x_{N}\right) $ is expressible in the form of
multidimensional Taylor series as follows

$%
\begin{array}{l}
f\left( x_{1},...,x_{N}\right) = \\
\\
\sum\limits_{l_{1}=0}^{\infty }...\sum\limits_{l_{N}=0}^{\infty }\frac{%
\left( -1\right) ^{l_{1}}}{l_{1}!}...\frac{\left( -1\right) ^{l_{N}}}{l_{N}!}%
\digamma \left( l_{1},...,l_{N}\right) \\
\\
\times
\;x_{1}^{a_{11}l_{1}+...+a_{1N}l_{N}+b_{1}}...%
\;x_{N}^{a_{N1}l_{1}+...+a_{NN}l_{N}+b_{N}}%
\end{array}%
$

\bigskip

then we obtain the expression

$%
\begin{array}{l}
J=\int\limits_{0}^{\infty }\frac{dx_{1}}{x_{1}}...\int\limits_{0}^{\infty }%
\frac{dx_{N}}{x_{N}} \\
\\
\sum\limits_{l_{1}=0}^{\infty }...\sum\limits_{l_{N}=0}^{\infty }\frac{%
\left( -1\right) ^{l_{1}}}{l_{1}!}...\frac{\left( -1\right) ^{l_{N}}}{l_{N}!}%
\digamma \left( l_{1},..,l_{N}\right) \\
\\
\times \;x_{1}^{a_{11}l_{1}+...+a_{1N}l_{N}+\widetilde{b}_{1}}..%
\;x_{N}^{a_{N1}l_{1}+...+a_{NN}l_{N}+\widetilde{b}_{N}}%
\end{array}%
$

\bigskip

being $\widetilde{b}_{i}=\nu _{i}+b_{i}$ $\left( i=1,...,N\right) $. After
applying sistematically Ramanujan's master theorem to the integral, we find
by method of induction, the general solution for this integral

\begin{equation}
\begin{array}{c}
J=\frac{1}{\left\vert \det \left( \mathbf{A}\right) \right\vert }\Gamma
\left( l_{1}^{\ast }\right) ...\Gamma \left( l_{N}^{\ast }\right) \digamma
\left( -l_{1}^{\ast },...,-l_{N}^{\ast }\right)%
\end{array}
\label{RMTG}
\end{equation}%
where $\det \left( \mathbf{A}\right) $ is evaluated by the formula

\begin{equation}
\det \left( \mathbf{A}\right) =\left\vert
\begin{array}{ccc}
a_{11} & \ldots & a_{1N} \\
\vdots & \ddots & \vdots \\
a_{N1} & \cdots & a_{NN}%
\end{array}%
\right\vert ,
\end{equation}%
and the variables $l_{i}^{\ast }\;\left( i=1,...,N\right) $ are solutions of
the following linear system

\begin{equation}
\left\{
\begin{array}{cc}
a_{11}l_{1}+...+a_{1N}l_{N}= & \widetilde{b}_{1} \\
\vdots & \vdots \\
a_{N1}l_{1}+...+a_{NN}l_{N}= & \widetilde{b}_{N}.%
\end{array}%
\right.
\end{equation}%
We have obtained Ramanujan's Master Theorem Generalized (RMTG). This general
formula $\left( \ref{RMTG}\right) $\ is equivalent to the general formula
obtained with the method of brackets for integral $J$ $\left( \ref{Brackets}%
\right) $. This result justifies mathematically the method of brackets as a
valid method for evaluating multidimensional integrals. Many examples are
discussed in references \cite{10,12}.

\section{Applications : Two detailed examples}

\subsection{Using Method of Brackets : Triangle diagram}

In the following we discuss two examples. The first example using the method
of brackets and a second example using RMTG. We start with the evaluation of
the following Feynman diagram

\begin{equation}
\begin{minipage}{3.0cm} \includegraphics[scale=.6]{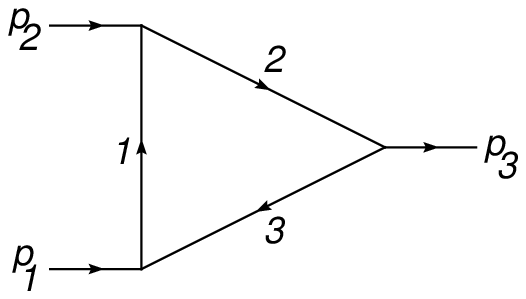}
\end{minipage}
\end{equation}%
The momentum integral for this graph is

\begin{equation}
\begin{array}{c}
G=\int \frac{d^{D}q}{i\pi ^{D/2}}\frac{1}{\left(
(p_{1}+q)^{2}-m_{1}^{2}\right) ^{a_{1}}} \\
\\
\times \frac{1}{\left( (p_{1}+p_{2}+q)^{2}-m_{2}^{2}\right)
^{a_{2}}(q^{2}-m_{3}^{2})^{a_{3}}}.%
\end{array}
\label{t7}
\end{equation}%
We also define the set $\left\{ a_{1},...,a_{N}\right\} $ as the set of
powers of the propagators, which in general can take arbitrary values. The
Schwinger's parametric\ representation of the Eq. $\left( \ref{t7}\right) $
is the following integral

\begin{equation}
\begin{array}{l}
G=\frac{(-1)^{-\frac{D}{2}}}{\Gamma (a_{1})\Gamma (a_{2})\Gamma (a_{3})}%
\int\limits_{0}^{\infty }d\overrightarrow{x}\;\exp \left(
\sum\limits_{k=1}^{3}x_{k}m_{k}^{2}\right) \\
\\
\times \frac{\exp \left( -\frac{C_{11}\;p_{1}^{2}+2C_{12}%
\;p_{1}.p_{2}+C_{22}\;p_{2}^{2}}{U}\right) }{U^{\frac{D}{2}}},%
\end{array}
\label{t1}
\end{equation}%
where $d\overrightarrow{x}=\prod\nolimits_{j=1}^{3}x_{j}^{a_{j}-1}dx_{j}$.
The polynomials $U=x_{1}+x_{2}+x_{3}$ and $C_{ij}$ are given by the
following equations

\begin{equation}
\begin{array}{l}
C_{11}=x_{3}(x_{1}+x_{2}), \\
\\
C_{12}=x_{2}x_{3}, \\
\\
C_{22}=x_{2}(x_{1}+x_{3}).%
\end{array}%
\end{equation}%
Inserting this in $\left( \ref{t1}\right) $ and remembering that $%
p_{3}^{2}=(p_{1}+p_{2})^{2}=p_{1}^{2}+2p_{1}.p_{2}+p_{2}^{2}$, after a
little algebra, we get the Schwinger's parametric representation of $\left( %
\ref{t7}\right) $
\begin{equation}
\begin{array}{l}
G=\frac{(-1)^{-D/2}}{\prod\limits_{j=1}^{3}\Gamma (a_{j})}%
\int\limits_{0}^{\infty }d\overrightarrow{x}\;\exp \left(
\sum\limits_{k=1}^{3}x_{k}m_{k}^{2}\right) \\
\\
\times \frac{\exp \left( -\frac{x_{3}x_{1}}{U}p_{1}^{2}\right) \exp \left( -%
\frac{x_{1}x_{2}}{U}p_{2}^{2}\right) \exp \left( -\frac{x_{2}x_{3}}{U}%
p_{3}^{2}\right) }{U^{D/2}}.%
\end{array}%
\end{equation}%
Now, we solve the diagram with the following conditions (for simplicity) : $%
m_{1}=m_{2}=0$, $m_{3}=M$, $p_{1}^{2}=p_{2}^{2}=0$ and $p_{3}=Q$, then, we
obtain the following integral for this case

\begin{equation}
\begin{array}{l}
G=\frac{(-1)^{-D/2}}{\prod\nolimits_{j=1}^{3}\Gamma (a_{j})}%
\int\limits_{0}^{\infty }d\overrightarrow{x}\;\exp \left( x_{3}M^{2}\right)
\\
\\
\times \frac{\exp \left( -\frac{x_{2}x_{3}}{U}Q^{2}\right) }{U^{D/2}}.%
\end{array}
\label{t2}
\end{equation}%
In the following we obtain the expansion of brackets (step by step). First,
we expand the exponential functions using rule $\left( I\right) $

\begin{itemize}
\item $\exp \left( x_{3}M^{2}\right) =\sum\limits_{n_{1}}\phi
_{n_{1}}\;\left( -M^{2}\right) ^{n_{1}}x_{3}^{n_{1}}.$

\item $\exp \left( -\frac{x_{2}x_{3}}{U}Q^{2}\right)
=\sum\limits_{n_{2}}\phi _{n_{2}}\;\left( Q^{2}\right) ^{n_{2}}\frac{%
x_{2}^{n_{2}}x_{3}^{n_{2}}}{U^{n_{2}}}.$
\end{itemize}

then, we obtain the integral

\begin{equation}
\begin{array}{ll}
G= & \frac{(-1)^{-D/2}}{\prod\nolimits_{j=1}^{3}\Gamma (a_{j})}%
\sum\limits_{n_{1},n_{2}}\phi _{n_{1},n_{2}}\ \left( Q^{2}\right) ^{n_{1}}
\\
&  \\
& \times \;\left( M^{2}\right) ^{n_{2}}\int\limits_{0}^{\infty }d%
\overrightarrow{x}\;\frac{x_{2}^{n_{2}}x_{3}^{n_{1}+n_{2}}}{U^{\frac{D}{2}%
+n_{2}}},%
\end{array}%
\end{equation}%
now, we expand the polynomial $U=\left( x_{1}+x_{2}+x_{3}\right) $ using
rule $\left( III\right) $

\begin{equation}
\begin{array}{l}
\frac{1}{\left( x_{1}+x_{2}+x_{3}\right) ^{\frac{D}{2}+n_{2}}}%
=\sum\limits_{n_{3},..,n_{5}}\phi _{n_{3},..,n_{5}}\; \\
\\
\times \;x_{1}^{n_{3}}x_{2}^{n_{4}}x_{3}^{n_{5}}\frac{\left\langle \frac{D}{2%
}+n_{2}+n_{3}+n_{4}+n_{5}\right\rangle }{\Gamma (\frac{D}{2}+n_{2})},%
\end{array}%
\end{equation}%
then using rule $\left( II\right) $ and a little algebra allows us to find
the expansion of brackets associated to the integral $\left( \ref{t2}\right)
$

\begin{equation}
\begin{array}{ll}
G= & \frac{(-1)^{-D/2}}{\prod\nolimits_{j=1}^{3}\Gamma (a_{j})}%
\sum\limits_{n_{1},..,n_{5}}\phi _{n_{1},..,n_{5}}\ \left( -M^{2}\right)
^{n_{1}} \\
&  \\
& \times \;\left( Q^{2}\right) ^{n_{2}}\frac{\prod\nolimits_{j=1}^{4}\Delta
_{j}}{\Gamma (\frac{D}{2}+n_{2})},%
\end{array}
\label{t3}
\end{equation}%
where the symbols $\left\{ \Delta _{j}\right\} $ represent the brackets

\begin{equation}
\left\{
\begin{array}{l}
\Delta _{1}=\left\langle \frac{D}{2}+n_{2}+n_{3}+n_{4}+n_{5}\right\rangle ,
\\
\Delta _{2}=\left\langle a_{1}+n_{3}\right\rangle , \\
\Delta _{3}=\left\langle a_{2}+n_{2}+n_{4}\right\rangle , \\
\Delta _{4}=\left\langle a_{3}+n_{1}+n_{2}+n_{5}\right\rangle .%
\end{array}%
\right.
\end{equation}%
Using rule $\left( IV\right) $ we find finally the solution to $\left( \ref%
{t2}\right) $. In this case exists two kinematical regions : $\left\vert
\frac{M^{2}}{Q^{2}}\right\vert <1$ and $\left\vert \frac{Q^{2}}{M^{2}}%
\right\vert <1$. The solution in the region $\left\vert \frac{M^{2}}{Q^{2}}%
\right\vert <1$\ is given by the following expression

\begin{equation}
\begin{array}{l}
G\left( \frac{M^{2}}{Q^{2}}\right) =\left( -1\right) ^{\frac{D}{2}}\left(
Q^{2}\right) ^{\frac{D}{2}-a_{123}} \\
\\
\times \;\frac{\Gamma \left( \frac{D}{2}-a_{12}\right) \Gamma \left( \frac{D%
}{2}-a_{13}\right) \Gamma \left( a_{123}-\frac{D}{2}\right) }{\Gamma \left(
a_{2}\right) \Gamma \left( a_{3}\right) \Gamma \left( D-a_{123}\right) } \\
\\
\times \ _{2}F_{1}\left( \left.
\begin{array}{c}
\begin{array}{cc}
1+a_{123}-D, & a_{123}-\frac{D}{2}%
\end{array}
\\
1+a_{13}-\frac{D}{2}%
\end{array}%
\right\vert \frac{M^{2}}{Q^{2}}\right) .%
\end{array}
\label{t8}
\end{equation}%
The solution for the region $\left\vert \frac{Q^{2}}{M^{2}}\right\vert <1$
are obtained by analytic continuation of $\left( \ref{t8}\right) $.

\subsection{Using RMTG : Massless bubble diagram}

We will solve the following diagram

\begin{equation}
\begin{minipage}{3.0cm} \includegraphics[scale=.6]{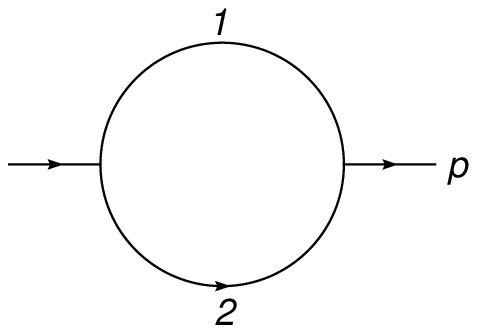} \end{minipage}
\label{t6}
\end{equation}%
the Schwinger's parametric representation is given by the expression

\begin{equation}
\begin{array}{l}
G=\frac{(-1)^{-\frac{D}{2}}}{\Gamma (a_{1})\Gamma (a_{2})}%
\int\limits_{0}^{\infty }\int\limits_{0}^{\infty
}dxdy\;x^{a_{1}-1}y^{a_{2}-1} \\
\\
\times \;\frac{\exp \left( -\frac{xy}{x+y}\;p^{2}\right) }{\left( x+y\right)
^{\frac{D}{2}}}.%
\end{array}%
\end{equation}%
Now, we expand the integrand using conventional mathematics

\begin{equation}
\begin{array}{c}
\exp \left( -\frac{xy}{x+y}\;p^{2}\right) =\sum\limits_{n=0}^{\infty }\frac{%
\left( -1\right) ^{n}}{n!}\left( p^{2}\right) ^{n}\frac{x^{n}y^{n}}{\left(
x+y\right) ^{n}},%
\end{array}%
\end{equation}%
resulting the following integral

\begin{equation}
\begin{array}{ll}
G= & \frac{(-1)^{-\frac{D}{2}}}{\Gamma (a_{1})\Gamma (a_{2})}%
\int\limits_{0}^{\infty }\int\limits_{0}^{\infty
}dxdy\;x^{a_{1}-1}y^{a_{2}-1} \\
&  \\
& \times \;\sum\limits_{n=0}^{\infty }\frac{\left( -1\right) ^{n}}{n!}\left(
p^{2}\right) ^{n}\frac{x^{n}y^{n}}{\left( x+y\right) ^{\frac{D}{2}+n}},%
\end{array}
\label{t5}
\end{equation}%
then, we expand the binomial in the denominator

\begin{equation}
\begin{array}{l}
\frac{1}{\left( x+y\right) ^{\frac{D}{2}+n}}= \\
\sum\limits_{k=0}^{\infty }\frac{\left( -1\right) ^{k}}{k!}\left( \frac{D}{2}%
+n\right) _{k}x^{-\frac{D}{2}-n-k}y^{k},%
\end{array}%
\end{equation}%
replacing in $\left( \ref{t5}\right) $ and doing a change of variables : $%
x\longrightarrow \frac{1}{x}$, we obtain finally the optimal structure for
applying RMTG, this is

\begin{equation}
\begin{array}{l}
G=\frac{(-1)^{-\frac{D}{2}}}{\Gamma (a_{1})\Gamma (a_{2})}%
\int\limits_{0}^{\infty }\int\limits_{0}^{\infty }dxdy\;x^{-a_{1}+\frac{D}{2}%
-1}y^{a_{2}-1} \\
\\
\times \sum\limits_{k=0}^{\infty }\sum\limits_{n=0}^{\infty }\frac{\left(
-1\right) ^{n}}{n!}\frac{\left( -1\right) ^{k}}{k!}\left( p^{2}\right) ^{n}
\\
\\
\times \left( \frac{D}{2}+n\right) _{k}x^{k}y^{k+n},%
\end{array}%
\end{equation}
allows us to obtain the solution of the diagram $\left( \ref{t6}\right) $

\begin{equation}
\begin{array}{ll}
G= & (-1)^{-\frac{D}{2}}\left( p^{2}\right) ^{\frac{D}{2}-a_{1}-a_{2}} \\
&  \\
& \times \frac{\Gamma (a_{1}+a_{2}-\frac{D}{2})\Gamma (\frac{D}{2}%
-a_{1})\Gamma \left( \frac{D}{2}-a_{2}\right) }{\Gamma (a_{1})\Gamma
(a_{2})\Gamma \left( D-a_{1}-a_{2}\right) }%
\end{array}%
\end{equation}

\section{Conclusions}

\qquad The method of brackets has been presented as an competitive
alternative compared with other advanced techniques for evaluating Feynman
diagrams. The main advantage of this technique is that it is systematic and
it does not require advanced mathematical tools, just linear algebra.
Although, this technique is not fully explained by RMTG, the results
obtained by method of brackets are identical to the obtained with RMTG when
the number of the summation indices is the same as the number of the
brackets. If the number of the brackets is less than the number of the
summation indices, RMTG is not useful to explain the obtained results by
applying method of brackets, although these results are correct. We are
currently working on studies to fully validate this technique through the
mathematical point of view.

\bigskip

\textbf{Acknowledgments}. The author would like to thank the organizers of
the conference Loops and Legs in Quantum Field Theory for the invitation and
for the financial support. For me is a survey to my work.

\end{document}